\documentclass[iop]{emulateapj}

\newcommand{\apjlet}{ApJL}

\shorttitle{The evolution of galaxies resolved in space and time: an inside-out growth view from the CALIFA survey}
\shortauthors{P\'erez et al.}

\begin{document}


\title{The evolution of galaxies resolved in space and time: an inside-out growth view from the CALIFA survey}


\author{E. P\'erez\altaffilmark{1} }
\author{R. Cid Fernandes\altaffilmark{1,2}}
\author{R. M. Gonz\'alez Delgado\altaffilmark{1}}
\author{R. Garc\'\i a-Benito\altaffilmark{1}}
\author{S. F. S\'anchez\altaffilmark{1,3}}
\author{B. Husemann\altaffilmark{4}}
\author{D. Mast\altaffilmark{1,3}}
\author{J.R. Rod\'on\altaffilmark{1}}
\author{D. Kupko\altaffilmark{4}}
\author{N. Backsmann\altaffilmark{4}}
\author{A.L. de Amorim\altaffilmark{2} }
\author{G. van de Ven\altaffilmark{5} }
\author{J. Walcher\altaffilmark{4} }
\author{L. Wisotzki\altaffilmark{4}}
\author{C. Cortijo-Ferrero\altaffilmark{1}}
\author{CALIFA collaboration\altaffilmark{6}}

\affil{\altaffilmark{1}Instituto de Astrof\'\i sica de Andaluc\'\i a (CSIC), Glorieta de la Astronom\'\i a s/n, E-18008 Granada, Spain.}
\affil{\altaffilmark{2}Departamento de F\'\i sica, Universidade Federal de Santa Catarina, P.O. Box 476, 88040-900, Florian—polis, SC, Brazil}
\affil{\altaffilmark{3}Centro Astron\'omico Hispano Alem\'an, Calar Alto, (CSIC-MPG), Jes\'us Durb\'an Rem\'on 2-2, E-04004 Almer\'\i a, Spain}
\affil{\altaffilmark{4}Leibniz-Institut fŸr Astrophysik Potsdam, An der Sternwarte 16, D-14482 Potsdam, Germany}
\affil{\altaffilmark{5}Max-Planck-Institut f\"ur Astronomie, K\"onigstuhl 17, D-69117 Heidelberg, Germany}
\affil{\altaffilmark{6}CALIFA International Collaboration \url{http://califa.caha.es}}

%Submitted to ApJL
\email{eperez@iaa.es}
\received{\em 2012 October 23}
\accepted{\em 2012 December 24}


%\altaffiltext{1}{Visiting Astronomer, Cerro Tololo Inter-American Observatory. CTIO is operated by AURA, Inc.\ under contract to the National Science Foundation.}


\begin{abstract}
The growth of galaxies is one of the key problems in understanding the structure and evolution of the universe and its constituents. Galaxies can grow their stellar mass by accretion of halo or intergalactic gas clouds, or by merging with smaller or similar mass galaxies. The gas available translates into a rate of star formation, which controls the generation of metals in the universe. The spatially resolved history of their stellar mass assembly has not been obtained so far for any given galaxy beyond the Local Group. Here we demonstrate how massive galaxies grow their stellar mass inside-out. We report the results from the analysis of the first 105 galaxies of the largest to date three-dimensional spectroscopic survey of galaxies in the local universe (CALIFA). We apply the fossil record method of stellar population spectral synthesis to recover the spatially and time resolved star formation history of each galaxy. We show, for the first time, that the signal of downsizing is spatially preserved, with both inner and outer regions growing faster for more massive galaxies. Further, we show that the relative growth rate of the spheroidal component, nucleus and inner galaxy, that happened 5--7 Gyr ago, shows a maximum at a critical stellar mass $\sim7\times10^{10}$ $M_{\sun}$. We also find that galaxies less massive than $\sim10^{10}$ $M_{\sun}$ show a transition to outside-in growth, thus connecting with results from resolved studies of the growth of low mass galaxies.
\end{abstract}





\section{Introduction}

The stellar mass of a galaxy, $M_\star$, is one of its most fundamental properties as it provides a
measure of the galaxy evolution process. Elucidating how it evolves in space and time, and
in relation to other galaxies is central to understanding the processes that govern the
efficiency and timing of star formation in galaxies. Look-back time studies of galaxy evolution
determine the spatially resolved star formation rate (SFR) at a
range of redshifts as a proxy for the average history of a particular type of galaxy
(Lilly et al. 1996; Madau et al. 1996; Folkes et al. 1999; York et al. 2000; Bell et al.
2004; Moles et al. 2008; Ilbert, et al. 2009; P\'erez-Gonz\'alez et al. 2008; Nelson et al.
2012). A major handicap of this approach is the inability to
follow the evolution of any one galaxy over time (Faber et al. 2007). In this work we follow
an alternative approach. For a new set of integral field spectroscopic (IFS) data of
a large sample of galaxies in the local universe, we recover their spatially
resolved individual star formation histories (SFH) by means of stellar population spectral synthesis. 

Beyond the Local Group,
where individual stars are not resolved, a powerful method to
reverse engineer the SFH of a galaxy is to find the most plausible combination of evolved
single stellar populations (SSP) that matches its spectrum (Walcher et al. 2011, and references therein). 
This fossil record method has been applied to images (e.g.
Brinchmann \& Ellis 2000; Bell \& de Jong 2000; P\'erez-Gonz\'alez et al. 2008; Hansson et al.
2012) with good spatial coverage but poor spectral resolution. It has also
been applied to Sloan Digital Sky Survey (SDSS) spectra of many galaxies
(Heavens et al. 2004; Brinchmann et al. 2004; Gallazzi et al. 2005; Cid Fernandes et al. 2005, 2011; 
Panter et al. 2007), although SDSS provides just a single spectrum 
per galaxy with varying spatial coverage with redshift. 
The CALIFA survey (S\'anchez et al. 2012) provides
a set of spectra that covers the extent of each galaxy with good
spatial and spectral sampling. Here we report on the
spatially resolved growth with cosmic time for 105 galaxies, 
which sample a wide range of properties in the color magnitude diagram (CMD),
drawn from the full CALIFA survey of 600 galaxies.

\begin{figure*}
\includegraphics*[width=0.8\textwidth,angle=90,trim=0 0 0 50 ,clip]{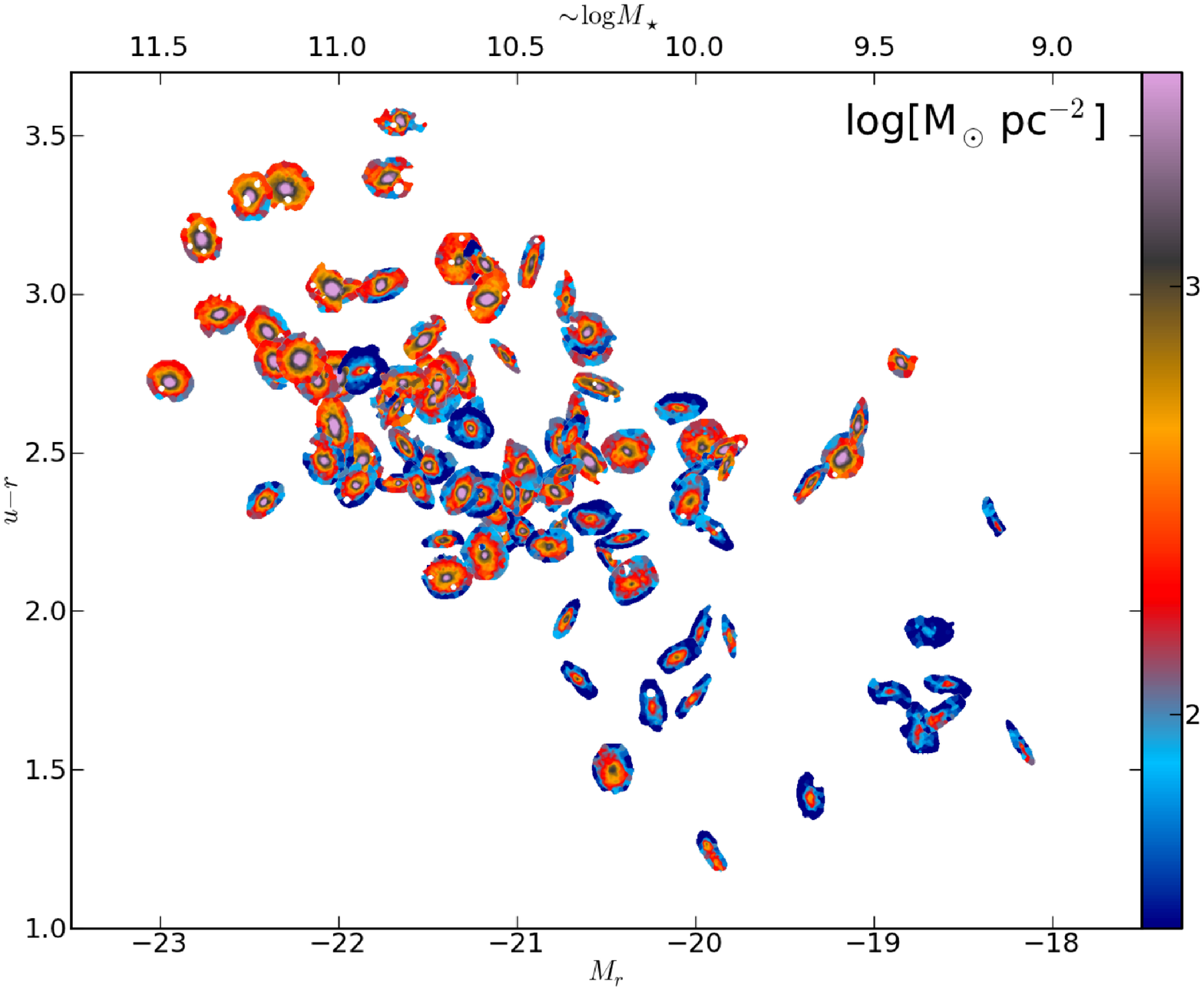}[ht!]
\caption{
Stellar mass surface density ($M_{\sun}$ pc$^{-2}$) in the galaxies CMD. 
Absolute magnitude in SDSS $r$, $M_r$, versus SDSS $u-r$ color, 
where each galaxy is represented with its present-day stellar mass density distribution. 
Holes in some galaxies result from removing foreground stars. 
The upper horizontal axis indicates approximate values of the stellar mass, 
from a linear fit to the $M_r$--log($M_\star$) relation. 
The time evolution of the stellar mass build up of each galaxy in this diagram can be seen in 
http://www.iaa.es/$\sim$eperez/CALIFA/CMD.
\label{fig1}}
\end{figure*}

\section{Sample and methods}

Details of the CALIFA survey, sample selection, 
reduction and calibration methods and analysis pipeline
are provided by S\'anchez et al. (2012), Husemann et al. (2012), 
Husemann et al. (2013, CALIFA first public data release DR1), Cid Fernandes et al. (2013),  and J. Walcher et al. (2013, in preparation).
Here we use the spectral datacubes combining both setups,
V500 and V1200 gratings, with median limiting sensitivity 1.0 and 2.2$\times10^{-18}$ 
erg s$^{-1}$ cm$^{2}$ \AA$^{-1}$ arcsec$^{-2}$,  spectral resolution $\sim6$ \AA\  in the range $\sim$3650--7500 \AA, 
median spatial resolution 3\farcs7, 
and spectrophotometric calibration better than 15\%.

For each galaxy the data are spatially rebinned to a minimum
signal-to-noise ratio of 20 in a continuum band at $5635\pm40$\AA\ 
(Cid Fernandes et al. 2012). 
The stellar spectral synthesis code {\sc starlight} 
(Cid Fernandes et al. 2005) is then used to recover the SFH,
including a value for the stellar extinction, from each spectrum. 
Here we use a set of SSPs sampling the 1Myr -- 14Gyr  age range with 40 age bins
and four metallicities between 0.2 and 2.5 solar (S. Charlot \& G. Bruzual 2007, private communication), assuming a Chabrier initial mass function. This method provides spatially resolved
SFHs for $\sim$200-2000 regions per galaxy, which we use to derive a three-dimensional buildup
of mass (with two spatial and one temporal coordinates). 

\section{Results}

Figure 1 shows the ($M_r$, $u-r$) CMD, where each galaxy is represented with its
present-day distribution of stellar mass density (in logarithmic units of $M_{\sun}$ pc$^{-2}$). Since
the absolute magnitude $M_r$ of galaxies is roughly proportional to their total log($M_\star$)
(a main driver of galaxy physics) and the $u-r$ color works as a rough evolutionary clock,
the CMD encapsulates galaxy evolution and the diversity of galaxies in a simple and
observationally convenient form. Furthermore, given its widespread use in the literature
(Faber et al. 2007; Blanton \& Moustakas 2009) and the fact the CALIFA sample was defined
to cover it as uniformly as possible, the CMD provides a natural framework to present and
analyze our results. The general trend in Figure 1 shows more luminous
galaxies to be redder, while the color coded mass density image of each galaxy
shows that more luminous ones have a higher central stellar surface density and steeper
gradients;\footnote{With few exceptions, such as the outstanding low surface brightness
galaxy at (-21.87, 2.76). } we now see these trends spatially resolved over a five magnitude range in
$M_r$. 

Spatially resolved mass distributions have been computed for some galaxies from
broad band imaging (e.g. Zibetti et al. 2009; Wuyts et al. 2012). Since we
have the spatially resolved SFH for each galaxy, we can also see the
time evolution of the CMD. This is shown in the animation
http://www.iaa.es/$\sim$eperez/CALIFA/CMD, with the 10000-250 Myr age plotted in the top
right-hand corner, stopping at a time when galaxies have most of their $M_\star$ in place. 
For clarity, and to emphasize the way 
$M_\star$ grows, galaxies are kept at their present location in the CMD, 
even though their luminosities and colors would actually
change with time. The most luminous, redder, massive galaxies have transformed an
important part of their mass into stars already 10 Gyr ago (e.g. P\'erez-Gonz\'alez et al. 2008),
particularly in their central regions, while many of the less massive galaxies show a
lower stellar surface density. 

Although we can derive the time evolution of the stellar populations at
their present day location, it is not feasible to trace their
spatial location at earlier epochs, i.e., we cannot follow the evolution of their motions.
If the stars undergo major spatial shuffling over their lives then our analysis would be hampered. 
Breaks in the exponential distribution of disk outer regions (e.g. Pohlen \& Trujillo 2006; Erwin et al. 2008)
have been interpreted as evidence of radial migration, although there is no general consensus.
Numerical simulations show that, for secular evolution, the consequences of radial migration are only significant
in the outer parts of galaxies, beyond about 2-3 scale lengths (Figure 2 in Ro\v{s}kar et al. 2008; Figures 15 and
16 in S\'anchez-Bl\'azquez et al. 2009; Behroozi et al. 2012). 
Here we partition galaxies into four radial zones to compare their histories,
so that the outer regions (at several effective radii) where radial migration may have an impact on the SFH
do not have a significant contribution to the results in the analysis that follows.

\begin{figure}
\includegraphics[width=0.5\textwidth,angle=0,trim=0 200 0 200,clip]{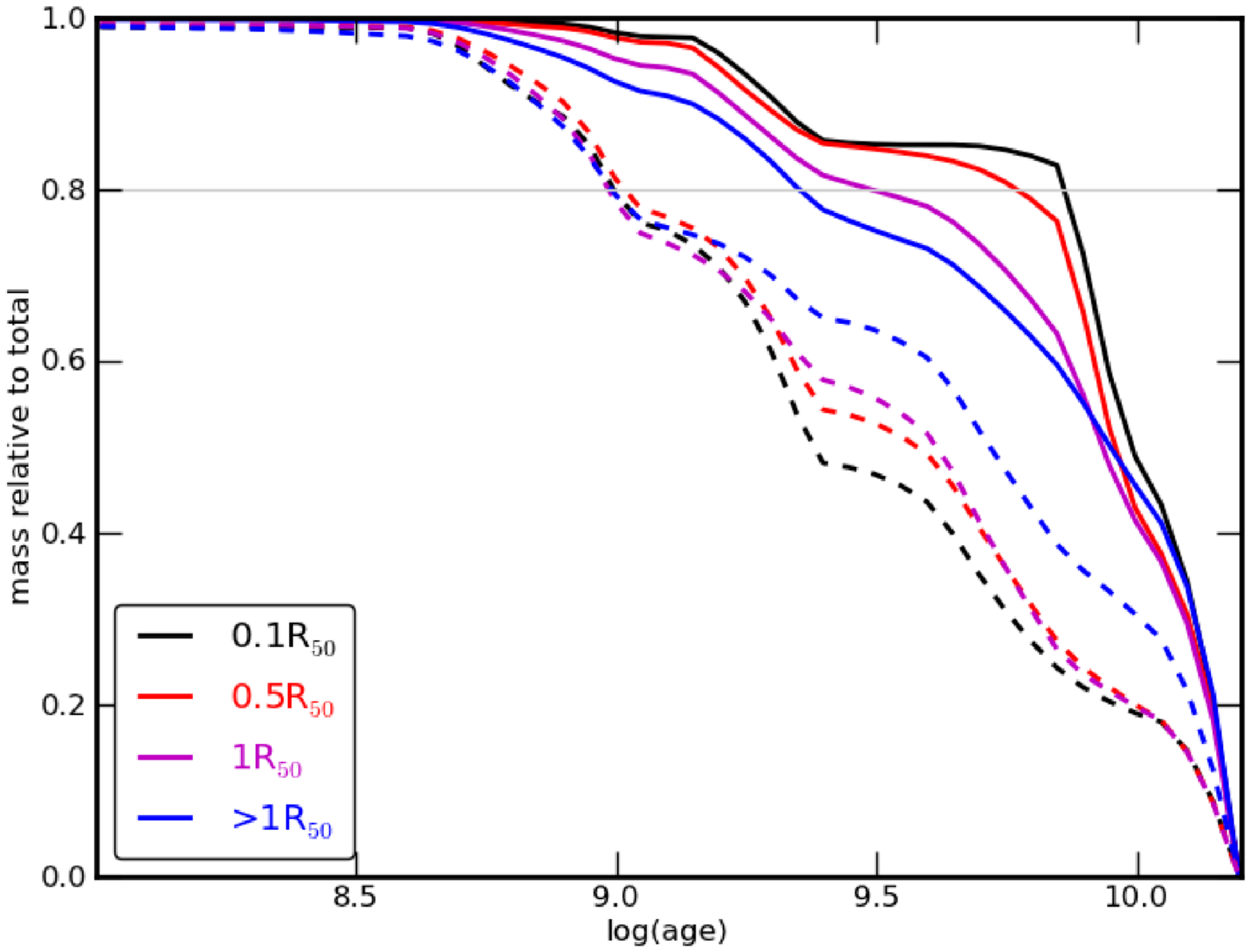}  % trim=left bottom right top
\caption{
Relative stellar mass growth of four main spatial components (nucleus, inner 0.5$R_{50}$ and 1$R_{50}$ and outer $>$1$R_{50}$) for two galaxy mass bins, the lowest (dashed, log $M_\star$ [$M_{\sun}$]$\sim9.58$) and an intermediate (full line, log $M_\star$ [$M_{\sun}$]$\sim10.83$). Each component is normalized to its final $M_\star$. To compute a growth rate, a horizontal gray line at 0.8 marks when each component achieves 80\% of its final $M_\star$. At the lowest mass, galaxies outskirts grow faster (dashed blue line steeper), but at higher masses the growth proceeds inside-out (full black line steeper).
\label{fig2}}
\end{figure}

\subsection{Spatially Resolved Growth of Stellar Mass $M_\star$}

The animated CMD is a visually compelling tool, but lacks the quantitative information needed to
probe the spatial growth of the stellar mass assembly. 
A more quantitative analysis in Figures 2-5 show
the time evolution for four galaxy regions: the nucleus (left, defined as
the central 0.1$R_{50}$\footnote{$R_{50}$ is the circular half light radius in $5635\pm40$\AA\ }), 
the region inside 0.5$R_{50}$, the region inside 1$R_{50}$,
and the region outside 1$R_{50}$. To make the growth of $M_\star$ more visible,
galaxies are normalized to a common radial scale in units of $R_{50}$, 
and then stacked in seven equally populated bins (15 galaxies each) of increasing
present-day galaxy $M_\star$. Figure 2 shows the relative growth of the four spatial
components for two $M_\star$ bins, the lowest (dashed, log $M_\star$ [$M_{\sun}$]$\sim$9.58) and an intermediate
mass (full line, log $M_\star$ [$M_{\sun}$]$\sim$10.83). To compute the $M_\star$ growth rate, a
gray line marks when each component achieves 80\% of its final $M_\star$. 
Clearly, at higher $M_\star$ the growth proceeds inside-out, while at the lowest $M_\star$ galaxies outskirts grow faster than their inner parts.

Figure 3 shows at a glance the trends for all mass bins. The horizontal axis represents log time 
(bottom) or redshift (top; $H_0=71$, $\Omega_{M}=0.27$, $\Omega_{\Lambda}=0.73$); 
the vertical axis represents the present day total $M_\star$. 
Panels show grayscale coded the relative growth of each of the four spatial components, normalized to its final
stellar mass, so that all components grow up to mass unity (i.e. grayscale
and contours in Figure 3 represent for all mass bins the growth curves shown for two example mass bins in  
Figure 2; the black contour line in each panel marks $0.8M_\star$). 
Thus, the relative growth of different parts of the galaxy can be
assessed. Three $M_\star$ growth cuts are plotted corresponding to three
mass bins (lowest, intermediate, and highest, in blue, green and red, as
colored in their axis mass value). The right-hand side panel shows that outer reaches
($>1R_{50}$) of less massive galaxies (log mass[$M_{\sun}$]$\sim$9.58, in the lower part of the plot)
achieve $0.8M_\star$ about 1 Gyr ago, while more massive ones (log
mass[$M_{\sun}$]$\sim$11.26) had $0.8M_\star$ already $\sim$5 Gyr ago. 
The systematics of this outer galaxy $M_\star$ growth is clearly seen from
the less massive up to the most massive galaxies in the sample. The central panels show a
similar behavior for the inner regions (1$R_{50}$ and 0.5$R_{50}$), but with a steeper
growth at higher galaxy masses. The left panel shows a similar
behavior for the nuclei, with an even steeper slope for galaxies more massive than
log mass[$M_{\sun}$]$\sim$10.5. Thus  there is a clear systematic
sequence of $M_\star$ growth from the inside out, where the mass is transformed into stars much earlier
within 0.5$R_{50}$.

\begin{figure*}
\includegraphics[width=0.47\textwidth,angle=90,trim=180 50 100 80,clip]{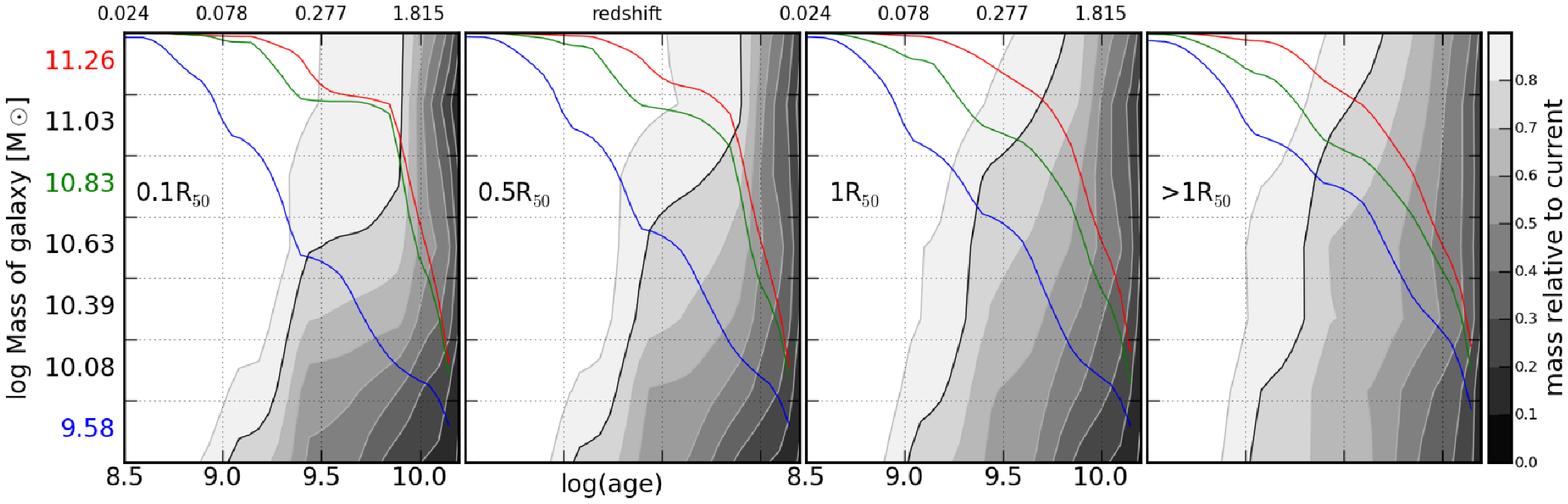}
\caption{
Inside-out stellar mass growth and spatially resolved downsizing. Panels show the normalized growth (grayscale in ten intervals of 10\%, with the 80\% contour in black) in time in the four zones of Figure 2. To avoid number statistics biases, the galaxies are stacked in seven equally populated bins of 15 galaxies each, sorted by present day total stellar mass; this implies that the vertical axis in figs. 3, 4, 5 is not uniform in log mass. For each spatial component the growth is normalized to its total final stellar mass (i.e. the growth curves of Figure 2 are cuts for two of the mass bins through the grayscale and contours). For each zone, three example mass bins growth curves are plotted (lowest, intermediate and highest, in blue, green and red) showing how the signal of downsizing is spatially preserved (the four blue lines here correspond to the four dashed lines in Figure 2, while the four green lines here correspond to the four full lines in Figure 2); the axis corresponding to these cuts is the same as the grayscale, i.e. the scale 0.0-1.0 on the right hand side. For a given galaxy mass, the progressive steepening of the growth from the outer ($>1$$R_{50}$) towards the nucleus shows the inside-out growth trend with the total mass of the galaxy. At the same time, for a given galaxy region, more massive galaxies grow faster than less massive ones, and this is sustained systematically from the inner to the outer zones.
}
\includegraphics[width=.6\textwidth,angle=90,trim=0 0 0 0 ,clip]{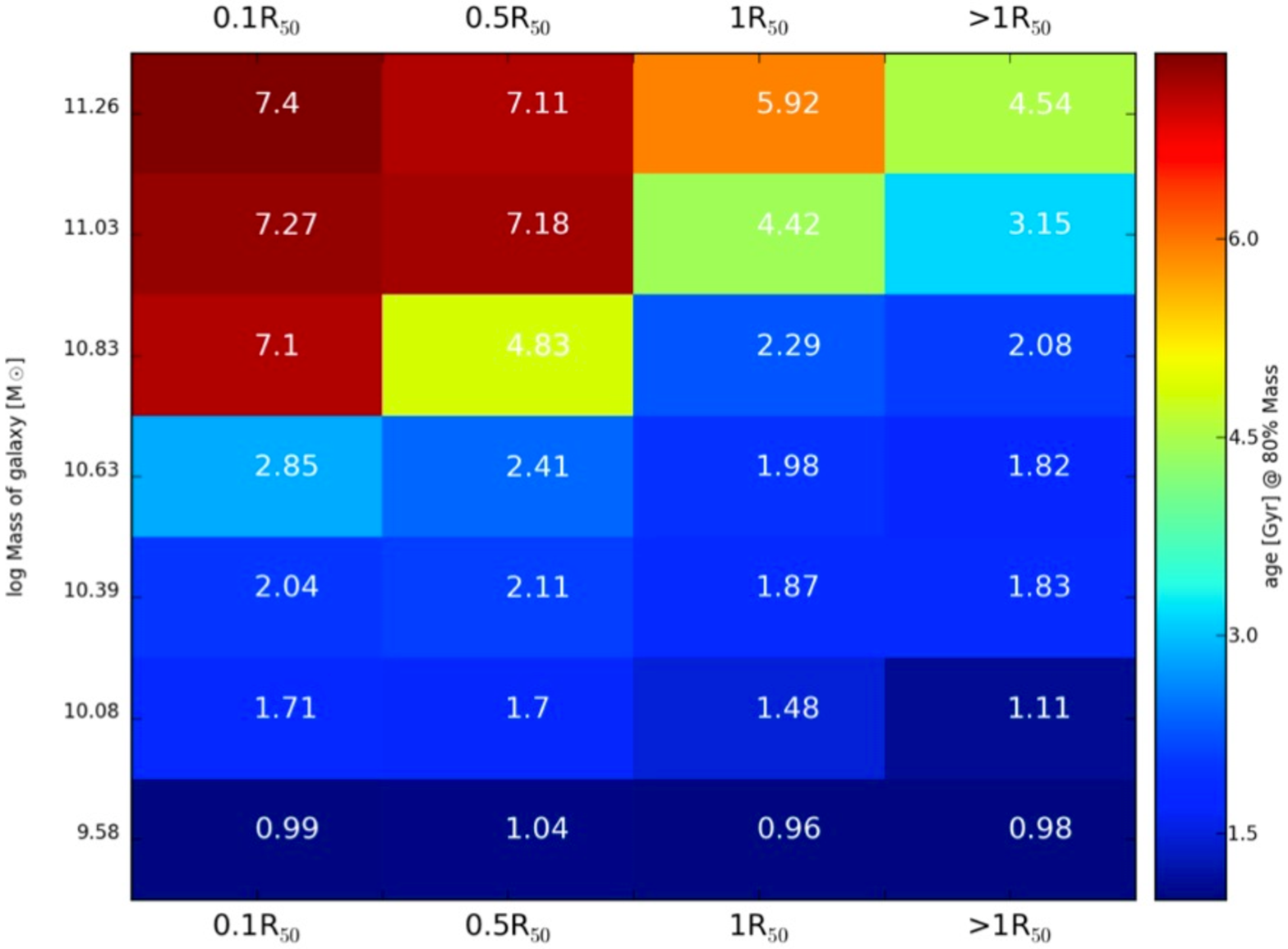}
\caption{
Inside-out stellar mass growth: ages. Using the same vertical axis as in Figure 3, this figure color codes the age at which each spatial component grows to 80\% of its final stellar mass; the age (in Gyr) is shown within each box. For a given mass, the horizontal run of color gives the systematic aging from the outer (right) to the inner regions (left). Conversely, for a given galaxy region, the vertical run gives the systematic aging of that zone with increasing galaxy total stellar mass. There is a systematic change both with mass and location for all cases, with the exception of the very low mass galaxies (bottom row) for which there is no inside-out growth (the age at 80\% growth is approximately the same in all four spatial components). These diagrams show clearly the differential inside-out growth and its explicit dependence on galaxy mass and galactocentric radius.
}
\end{figure*}

\subsection{Spatially Resolved Mean Age of the Stellar Populations}

To better understand this result, for each of the spatial regions we
compute the age at $0.8M_\star$ (as indicated in Figures 2 and 3). 
These are represented color coded in Figure 4 (in Gyr, also shown in each box), 
with the present  day stellar mass of the galaxy in
the vertical axis and the spatial zone in the horizontal. 
Arising from the number of galaxies, statistical errors in the age for each box amount to 0.23dex,
while systematic errors introduced by the choice of other SSP basis (Gonz\'alez Delgado et al. 2005) are 0.05dex
(R. M. Gonz\'alez Delgado et al. 2013, in preparation).
At a glance it is clear that: 
(1) running vertically for a given galaxy zone, older ages (red) correspond to more massive galaxies, 
and 
(2) running horizontally for a given galaxy mass, older ages (red) correspond to the central zones. 
The trends are clear for all masses and spatial regions, except for the lower mass bin, 
where there is no evidence of inside-out growth. 
In fact, the evidence at the lower masses is that galaxies do
grow outside-in (Figure 2), connecting with the behavior seen
in studies of dwarf galaxies (e.g. Gallart et al. 2008; Zhang et al.
2012, and references therein).

\begin{figure*}
\includegraphics[width=.5\textwidth,angle=90,trim=100 50 110 50 ,clip]{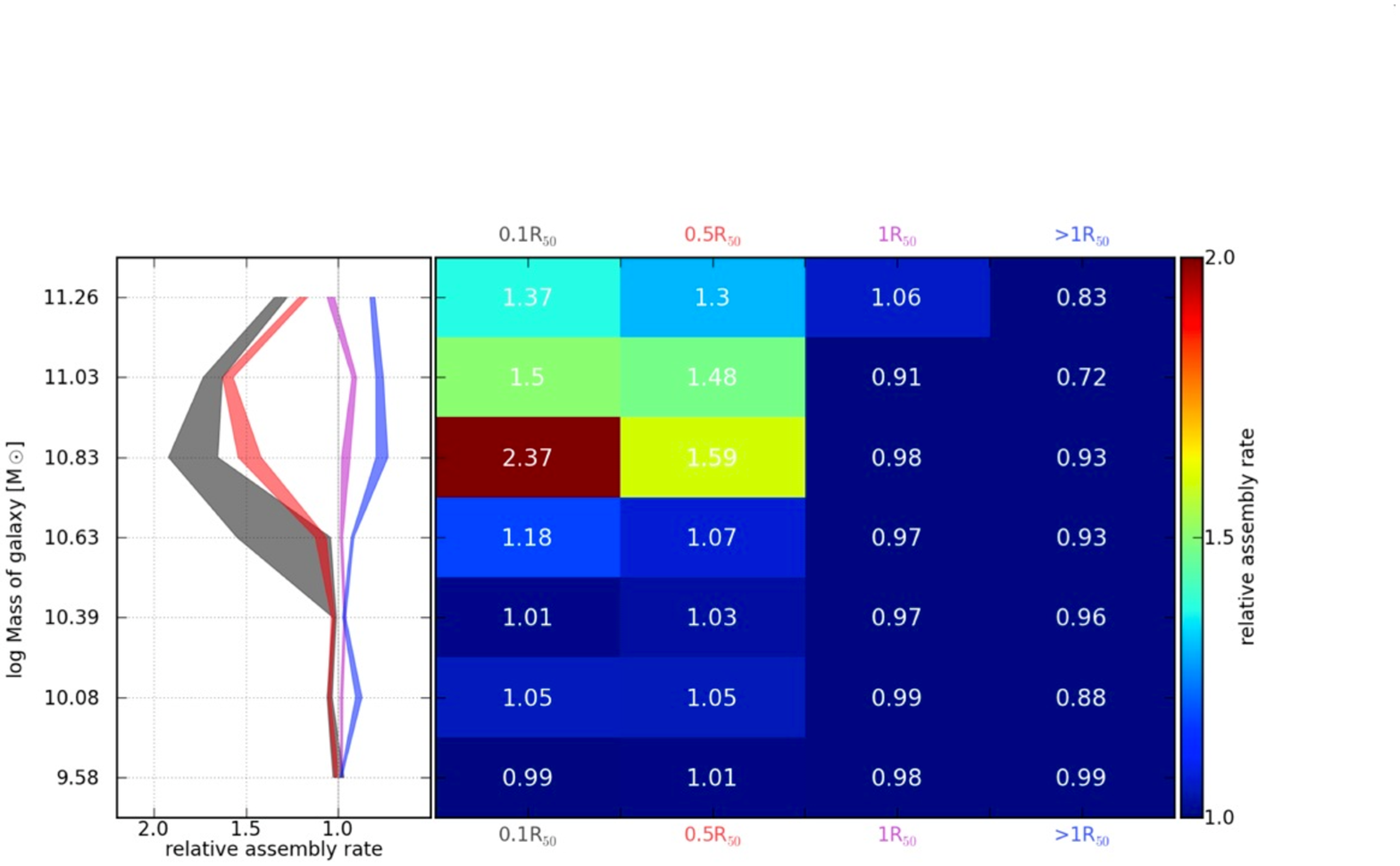}
\caption{
Stellar mass assembly rate. With the same axes of Figure 4, the right panel color codes the relative growth rate at 80\% mass, for a given galaxy mass and zone, normalized to the total galaxy growth rate at 80\%. The result indicates that for most of the galaxies, on average, the relative rate (within each galaxy) of transforming gas into stars is similar in all parts, except for the central regions ($<$0.5$R_{50}$) of galaxies more massive than $\sim5\times10^{10}$ $M_{\sun}$, which are up to a factor of $\sim2$ faster. The left panel shows vertical cuts through the four zones, including uncertainty bands computed in 1000 iterations of randomly removing five different galaxies at a time.
}
\end{figure*}

The increase in mean stellar ages with increasing $M_\star$ reflects the downsizing phenomenon 
previously established on the basis of fossil record analyses of spatially unresolved data 
(Heavens et al. 2004; P\'erez-Gonz\'alez et al. 2008). Figures 3 and 4 show, for the first time, that the signal of downsizing is preserved in a radial fashion, with both inner and outer regions growing faster
for more massive galaxies. At the same time, these figures show that different spatial
regions in a galaxy grow at different paces.

\subsection{Spatially Resolved Growth Rate of the Stellar Populations}

To examine the relative growth of inner and
outer regions more closely, we compute growth rates from the inverse of the cosmic time
span corresponding to the build up of $0.8M_\star$. This is done for each
spatial component of each galaxy, and the resulting rates expressed in
units of the growth rate of the galaxy as a whole, obtained by integrating over its full
spatial extent. Thus we derive the relative growth rates shown in Figure 5. 
The right panel, with coordinates as in Figure 4, codes in
color this relative growth rate. The left panel plots vertical cuts through the four
locations within the galaxy, as a function of total present day stellar mass.
This figure shows at once that not only the inner parts of more massive galaxies grew $M_\star$
faster (as in Figure 4), but also that in galaxies more massive than $5\times10^{10}$ $M_{\sun}$ the
inner regions grew as much as 50\%-100\% faster than the galaxy-wide average. The relative
growth rate for 1$R_{50}$ (magenta line) is close to unity, meaning that $R_{50}$
best represents the stellar mass growth in the galaxy as a whole.

Although we have grouped 105 galaxies (105123 spectra in total) in just
a few mass bins to obtain robust results, when the complete CALIFA sample of 600 galaxies
is observed and analyzed these results will be established with a finer detail. The most
striking aspect of Figure 5 is the peak at $M_\star\sim6-7\times10^{10}$ $M_{\sun}$. At masses below this
the relative growth rates of inner and outer regions slowly converge to the galaxy average,
becoming essentially indistinguishable at the lowest galaxy masses sampled. A convergence towards
spatially uniform histories is also seen at masses above the critical mass, but
the inner regions still grow faster than the outer ones at the highest masses. 
The largest differences peak at $\sim6-7\times10^{10}$ $M_{\sun}$, where the nucleus
grows about twice faster than the outer regions. 

\section{Discussion and summary}

Previous studies have revealed a special mass or mass range (referred to as critical or pivotal) around $M_\star\sim6\times10^{10}$ $M_{\sun}$ 
(Kauffmann et al. 2003; Mateus et al. 2006; Leauthaud et al. 2012, and references therein). 
Leauthaud et al study the stellar-to-halo mass relation (SHMR) using COSMOS data and find that this mass marks where 
Õthe accumulated stellar growth of the central galaxy has been the most efficientÕ. 
On the modeling side, Shankar et al. (2006) discuss the shape of the SHMR as a change of dominance of two feedback mechanisms, 
with stellar winds and supernova (SN) efficiently removing the gas and quenching the star formation at lower galaxy masses, 
while active
galactic nucleus (AGN) winds become more powerful in massive galaxies thus lowering star formation efficiency at higher masses. 
There is still much to be understood about the self-consistent combined effects of AGN and SN feedback (Stinson et al. 2013; Booth \& Schaye 2012). 
Recent papers try to reproduce the observed SHMR from cosmological simulations (cf. Figure 14 in Behroozi et al. 2012), 
with values of $\sim2-10\times10^{10}$ $M_{\sun}$ in the peak efficiency of conversion of halo gas into stellar mass. 
Cosmological models of star formation with negative feedback do not allow yet for a more precise value of this critical mass for optimal growth (De Lucia \& Borgani 2012). 

In short, there is clear mounting evidence from a series of independent studies, both observationally and from numerical simulations, 
of a special galaxy mass range where the SFR reaches higher values faster. 
Our results, Figure 5, show that the physical process(es) responsible for this peak efficiency also leave a strong imprint in the {\em spatial} assembly of stars in galaxies. 
From our spatially resolved study we can see that this is restricted to the inner reaches of galaxies with this pivotal mass. 
This would imply that the effect is seen in the spheroidal component, which dominates the central mass distribution of galaxies at this stellar mass range, 
and that this happened 5-7 Gyr ago (a redshift $\sim$0.5--1; Figure 4).

We can also tentatively frame our results in the context of secular versus merger driven evolution. 
A possible scenario to interpret our findings is that galaxies with $M_\star\gtrsim5\times10^{10}$ $M_{\sun}$ 
have grown quickly their inner part by means of a merger at ages 5-9 Gyr ago (such as M31), 
while lower mass galaxies are dominated by secular evolution (such as the Milky Way). 
Although we do not have the means to test this, it is compatible with the results obtained by Puech et al. (2012), 
and with the studies of M31 and the Milky Way (Hammer et al. 2007, 2010; Cignoni et al. 2006; Vergely et al. 2002; McMillan 2011).

Our analysis of the SFH in 105 galaxies of the CALIFA survey provides a direct powerful probe of the rates and efficiency 
of spatially resolved stellar mass growth in galaxies,  as derived from the information encoded in their present-day spectra.
We demonstrate spatially resolved downsizing, with galaxies more massive than M$_\star\sim5\times10^{9}$ $M_{\sun}$ 
growing inside-out while lower mass galaxies grow outside-in. We find an optimal growth for the galaxy stellar mass around $\sim7\times10^{10}$ $M_{\sun}$.


%% Included in this acknowledgments section are examples of the
%% AASTeX hypertext markup commands. Use \url without the optional [HREF]
%% argument when you want to print the url directly in the text. Otherwise,
%% use either \url or \anchor, with the HREF as the first argument and the
%% text to be printed in the second.

\acknowledgments

This Letter is based on data obtained by the CALIFA survey (\url{http://califa.caha.es}), 
funded by the Spanish MINECO grants ICTS-2009-10, AYA2010-15081, 
and the CAHA operated jointly by the Max-Planck IfA and the IAA (CSIC). 
We benefited from discussions during the CALIFA busyweeks. 
The CALIFA Collaboration also thanks the CAHA staff for the dedication to this project.

\end{document}